# DisastRAG: A Multi-Source Disaster Information Integration and Access System Based on Retrieval-Augmented Large Language Models

Bo Li, Zhitong Chen, Kai Yin, Junwei Ma, Yiming Xiao, Ali Mostafavi

UrbanResilience.AI Lab, Zachry Department of Civil and Environmental Engineering, Texas A&M University, College Station, Texas, United States.

## Abstract

Effective disaster management requires rapid access to information distributed across structured operational records, unstructured institutional documents, and dynamic external sources. However, most existing disaster information systems and retrieval-augmented generation frameworks remain organized around a single access pathway, limiting their ability to support heterogeneous, time-sensitive, and context-dependent information needs. This study presents DisastRAG, a disaster-aware information integration and access system that combines large language models with retrieval-augmented access to structured, unstructured, and contextual disaster information. The framework is built around a multi-path architecture that supports document retrieval over a curated hazard corpus, structured access over relational disaster records, and external web fallback for out-of-corpus requests, while also incorporating query understanding, strategy routing, response generation, and contextual memory within a unified system. We evaluated the document retrieval performance using four open-source large language models across multiple retrieval configurations on multiple-choice and open-ended disaster information tasks. Retrieval augmentation consistently improves performance over no-retrieval baselines, yielding multiple-choice gains of 12–23 percentage points and open-ended keypoint coverage gains of up to 10.5 percentage points. Results show that larger candidate pools are most helpful for weaker models, while stronger models are more sensitive to retrieval noise. Hybrid retrieval performs best for open-ended coverage, whereas vector retrieval and shallower reranking more often favor closed-form factual selection. Case studies further show that structured access and web fallback extend the framework beyond document-only RAG.

# 1. Introduction

Disaster management increasingly depends on the ability to access, synthesize, and interpret information from multiple heterogeneous sources under severe time constraints (Fan, Zhang et al. 2021, Camps-Valls, Fernández-Torres et al. 2025). During an unfolding event, decision-makers must move rapidly across structured records (e.g., power outage records, shelter capacity, evacuation statistics, and exposure indicators) as well as unstructured materials (e.g., emergency manuals, hazard assessments, incident reports, and operational advisories) (Li, Ma et al. 2025). Although these resources are often available in institutional repositories, they are rarely organized as a unified operational information system (Lee, Wang et al. 2012). Disaster knowledge is typically fragmented across incompatible databases, static reports, and isolated document collections, creating a persistent gap between the existence of information and its effective use in time-sensitive decision support (De Bartolomeo and De Nicola 2025, Sarker 2025).

This fragmentation is not merely a data management problem; it is fundamentally a system integration problem. Traditional disaster information platforms are typically designed around a single modality or a single access paradigm (Li, Xie et al. 2017, Lei, Dong et al. 2025, Ma, Li et al. 2025). Dashboard-based systems can summarize structured indicators efficiently, but they provide limited support for navigating procedural guidance, explanatory documents, or policy-oriented text (U.S. Federal

Emergency Management Agency 2025, PowerOutage.us 2026, U.S. National Weather Service 2026). In contrast, document repositories contain rich contextual and institutional knowledge yet are poorly suited for rapid, interactive querying during emergencies. As a result, responders and analysts are often forced to manually coordinate across multiple tools, interfaces, and data formats in order to answer even moderately complex questions, thereby degrading both the speed and reliability of situational awareness.

Recent advances in large language models (LLMs) and retrieval-augmented generation (RAG) have created new opportunities for rethinking how disaster information systems are designed and accessed. LLMs are particularly well suited for disaster-management settings because they enable users to express complex information needs in natural language and receive coherent, context-sensitive responses without requiring specialized knowledge of database schemas, dashboard logic, or document organization (Yin, Li et al. 2024, Liu, Yin et al. 2025, Chen, Yin et al. 2026). This capability is especially valuable in time-critical environments, where responders and analysts must rapidly interpret diverse forms of information and translate them into actionable understanding. However, despite these strengths, LLMs used in isolation remain limited as operational decision-support tools. Their responses are generated primarily from parametric memory, which may be incomplete, outdated, or unverifiable, and they are therefore vulnerable to hallucination and source inconsistency in high-stakes settings (Farquhar, Kossen et al. 2024). Retrieval-augmented generation (RAG) addresses this limitation by coupling language generation with external evidence retrieval (Gao, Xiong et al. 2023). By combining natural-language interaction with access to authoritative data and documents, RAG makes it possible to generate responses that are not only fluent and contextually relevant but also grounded in traceable sources rather than relying solely on model memory (Lewis, Perez et al. 2020). In principle, such a system could allow users to query disaster information conversationally, integrate evidence across data modalities, and receive responses that are both interpretable and operationally useful.

However, current RAG implementations remain limited in three ways that are particularly consequential for disaster-oriented applications. First, most existing systems are commonly designed around a single retrieval path over unstructured text. Zhang et al. (2025) noted that conventional RAG architectures, which are primarily optimized for unstructured textual corpora, encounter substantial difficulties when extended to enterprise environments containing structured, semi-structured, and tabular data (Yan, Zhang et al. 2025). In disaster settings, many disaster questions require access to structured records, narrative documents, and sometimes external sources within the same workflow (Sagun, Bouchlaghem et al. 2009, Chen, Yin et al. 2026). For example, a query about peak county-level outages during Hurricane

Harvey requires access to structured records, while a follow-up about evacuation protocols in those areas demands document retrieval. Such heterogeneous information needs are not well supported by single-path RAG systems. Second, many RAG systems are still implemented as one-shot question-answering pipelines rather than as coordinated information systems capable of maintaining interaction history, supporting iterative follow-up queries, and dynamically adapting retrieval behavior to different categories of user requests (Singh, Ehtesham et al. 2025). This architectural rigidity limits their usefulness in disaster response contexts, where information seeking is often incremental, context-dependent, and shaped by evolving operational needs (Dong, Mi et al. 2026). In practice, decision-makers rarely ask isolated questions; instead, they engage in progressive inquiry, refining earlier requests, comparing evidence across sources, and adjusting their focus as the situation develops. Third, although retrieval performance is widely recognized as a foundational determinant of grounded generation quality, limited research has systematically examined how specific retrieval design choices, including retrieval strategy, candidate-pool construction, and reranking configuration, jointly influence downstream performance in disaster-domain information-seeking tasks (Han, Liu et al. 2025, Yin, Dong et al. 2025, Cui, Zhai et al. 2026). As a result, important questions remain about how to design retrieval pipelines that are both technically effective and operationally appropriate for high-stakes emergency settings.

To address these challenges, we propose DisastRAG, a disaster-aware information integration and access system that combined LLM-based generation with retrieval-augmented access to structured, unstructured, and dynamic external disaster information. The system integrates query understanding, strategy routing, multiple evidence pathways, response generation, and contextual memory within a unified architecture. Its goal is to support the end-to-end workflow of disaster information seeking, from interpreting a user's request and determining where relevant evidence most likely resides to retrieving and refining that evidence, generating a grounded response, and preserving interaction context for subsequent turns. The proposed system operates over a two-layer knowledge foundation. The structured layer contains tabular disaster-related records that support field-based and quantitative access, while the unstructured layer contains a curated corpus of hazard-related documents converted into machine-readable text for retrieval. Built on this foundation, the framework coordinates evidence acquisition and refinement prior to response generation, satisfying the heterogeneous information needs of disaster management. A central feature of the system is its multi-path architecture. Rather than forcing all requests through a single retrieval pipeline, the system routes queries across different evidence-access pathways according to their informational characteristics. Narrative or procedural requests are handled through document retrieval. Record-oriented or quantitative requests are mapped to structured access.

Out-of-corpus or time-sensitive requests are supported through external fallback. In addition, the framework incorporates system-level modules for conversational continuity and broader operational coverage, enabling more realistic interaction patterns. To assess the performance of the proposed system, this study systematically evaluates keyword-based, vector-based, and hybrid retrieval strategies alongside candidate-pool and reranking configurations, using both multiple-choice and open-ended task formats. To complement this quantitative evaluation, representative case studies illustrate how the broader system processes different classes of requests across pathways.

The remainder of this paper is organized as follows. Section 2 reviews related work on disaster information systems, retrieval-augmented generation, and large language models, and identifies the limitations of existing single-path approaches that motivate this study. Section 3 presents the architecture of DisastRAG, including the knowledge foundation, query understanding and strategy routing, evidence-access pathways, response generation, and contextual memory support. Section 4 describes the evaluation design, including the benchmark tasks, retrieval configurations, model backbones, and evaluation metrics. Section 5 reports quantitative results on multiple-choice and open-ended tasks across retrieval strategies and configurations. Section 6 presents case studies illustrating how the structured access and external fallback pathways support operational request types beyond the scope of document retrieval. Section 7 concludes with key findings, implications for retrieval system design, and directions for future work.

## 2. Related Work

### 2.1 Disaster Informatics and Disaster Information Systems

Disaster informatics refers to the study of how information is generated, organized, transmitted, understood, and used throughout the disaster management lifecycle (Yang, Zhang et al. 2020). Prior research has established that disaster information environments are inherently complex, as they involve multiple information sources, rapid temporal change, strong spatial dependence, distributed organizational responsibilities, and frequent uncertainty or inconsistency under real-world conditions (Hristidis, Chen et al. 2010). This complexity is closely related to the multi-phase nature of disaster management, which encompasses preparedness, response, and recovery, each associated with distinct information producers, users, urgency levels, and data modalities (Du, Mugambi et al. 2026).

Computational support for disaster information management has evolved from early GIS-based damage assessment and spatial decision support systems to more sophisticated monitoring portals, incident reporting platforms, and data-driven dashboards (Li, Xie et al. 2017). Modern platforms such as FEMA's operational databases, NOAA's hazard monitoring infrastructure, and utility-sector outage tracking systems have substantially expanded the availability of structured disaster data (U.S. Federal Emergency Management Agency 2025, PowerOutage.us 2026, U.S. National Weather Service 2026). However, these platforms remain largely query-inaccessible to non-technical users, and their utility under time pressure depends on analyst expertise in database querying, GIS operations, and cross-system data integration (Lei, Dong et al. 2025). The result is that data-rich institutions frequently depend on manual synthesis during active operations, introducing both delay and error.

More recent studies have explored machine learning and natural language processing (NLP) as means of reducing these barriers. For example, research in crisis informatics has demonstrated the value of automated classification, filtering, and information extraction for identifying disaster-relevant content from high-volume textual sources such as social media and emergency communications (Imran, Castillo et al. 2015, Zhang, Fan et al. 2019, Upadhyay, Meena et al. 2024). Related work has also examined the use of NLP techniques for event detection, situational summarization, and the identification of actionable information from unstructured disaster-related text (Fan, Jiang et al. 2020, Chen and Ji 2021, Upadhyay, Meena et al. 2024). These developments suggest that data-driven language technologies can improve the accessibility of disaster information by helping transform large volumes of dispersed textual content into more usable operational signals.

Nevertheless, most existing systems remain focused on a single data source or a narrow task, such as classification or extraction, rather than supporting unified access across structured operational records and unstructured document collections.

In addition, they generally provide limited support for context-preserving interaction and do not systematically examine how different information-access strategies affect downstream performance in disaster-oriented settings. These limitations motivate the need for more integrated disaster information access frameworks.

### 2.2 LLM-Enabled Retrieval-Augmented Information Access

The emergence of large language models has substantially reshaped the paradigm of information access by introducing natural language as a primary interface to complex knowledge environments (Breuer, Frihat et al. 2025). In contrast to conventional query systems that often require familiarity with database schemas, retrieval logic, or platform-specific interfaces, LLM-based systems allow users to express information needs in flexible natural language, thereby lowering technical barriers and enabling more exploratory forms of information seeking (Chen, Yuan et al. 2024). More fundamentally, LLMs can serve as reasoning layers that interpret user intent, connect dispersed pieces of evidence, and produce coherent and interpretable responses from heterogeneous sources (Shah, White et al. 2025).

Retrieval-augmented generation further advanced this paradigm by coupling language generation with external evidence retrieval, thereby enabling responses to be grounded in authoritative sources rather than relying solely on parametric memory (Zheng, Zhang et al. 2026). This architectural shift is particularly important in knowledge-intensive and high-stakes settings, where factual reliability, traceability, and source alignment are essential. Recent studies in disaster management have begun to extend LLM-based information-access systems through RAG. For example, Lei, Dong et al. (2025) document the growing use of LLMs across the disaster management lifecycle, including applications in which RAG pipelines are used to retrieve historical flood information and support risk-level recommendations (Lei, Dong et al. 2025). Xie, Jiang et al. (2025) introduce WildfireGPT, a RAG-based multi-agent system for wildfire resilience, and show that domain-specific retrieval augmentation can improve the contextual accuracy of LLM responses in hazard-focused settings (Xie, Jiang et al. 2025). Jiao, Park et al. (2025) similarly presents a modular RAG-based agent for emergency preparedness that provides grounded procedural guidance for emergency response scenarios (Jiao, Park et al. 2025).

In parallel, researchers have increasingly recognized that many information-access tasks cannot be adequately supported through text retrieval alone. Structured information access has therefore been explored through integrations with Text-to-SQL and related methods. TableRAG demonstrates that SQL-oriented reasoning over tabular data can substantially outperform document-based approximations when the target information is inherently structured (Yu, Jian et al. 2025). More recently, systems such as HetaRAG (Yan, Zhang et al. 2025) and the multi-agent RAG framework proposed by (Salve, Attar et al. 2024) have extended this line of work by routing queries across heterogeneous data stores, including vector indices, relational databases, and knowledge graphs, through modality-specific retrieval components or agents.

These studies suggest that recent RAG research is moving beyond single-corpus document retrieval toward more adaptive and heterogeneous information-access architectures (Evchina, Puttonen et al. 2015, Qu, Chen et al. 2025, Shi, Li et al. 2025). However, important limitations remain in disaster-management settings, where information needs are spatially anchored, temporally dynamic, and often require multi-turn interaction across both structured impact databases and curated hazard-related document collections under conditions of partial and evolving information availability. Existing disaster-oriented information-access systems rely heavily on static, single-step retrieval workflows, to emphasize text-only evidence access, and to provide only limited mechanisms for managing knowledge boundaries when local sources are insufficient. In addition, the relationship between retrieval design choices, including retrieval strategy, candidate-pool size, and reranking depth, and downstream information-access performance has not been systematically examined in this context.

# 3. System Architecture

## 3.1 Overview

The proposed system is a disaster-aware framework for information integration and access, built upon an LLM-powered, retrieval-augmented architecture. The framework coordinates five functional components: (1) multi-source knowledge foundation, (2) query understanding and strategy routing, (3) evidence-access pathways, (4) response generation, and (5) contextual memory support. It is designed to support the end-to-end operational workflow of information integration and access in disaster settings. Incoming user requests are first interpreted by the query understanding module, which extracts structured representations of the query. Then the representation is then passed to the strategy router, which dispatches the request to the most appropriate evidence-access pathway: the document retrieval branch for narrative, explanatory, and procedural questions; the structured access branch for quantitative or field-indexed requests over relational disaster records; or the external web fallback branch for requests that exceed the topical or temporal coverage of the internal knowledge base. Within these pathways, evidence is retrieved and being passed to the response generation module, which produces a grounded natural-language answer. A contextual memory component maintains recent interaction history and supplies relevant prior context to support follow-up queries and coherent multi-turn interaction. Figure 1 provides a schematic overview of this end-to-end architecture.

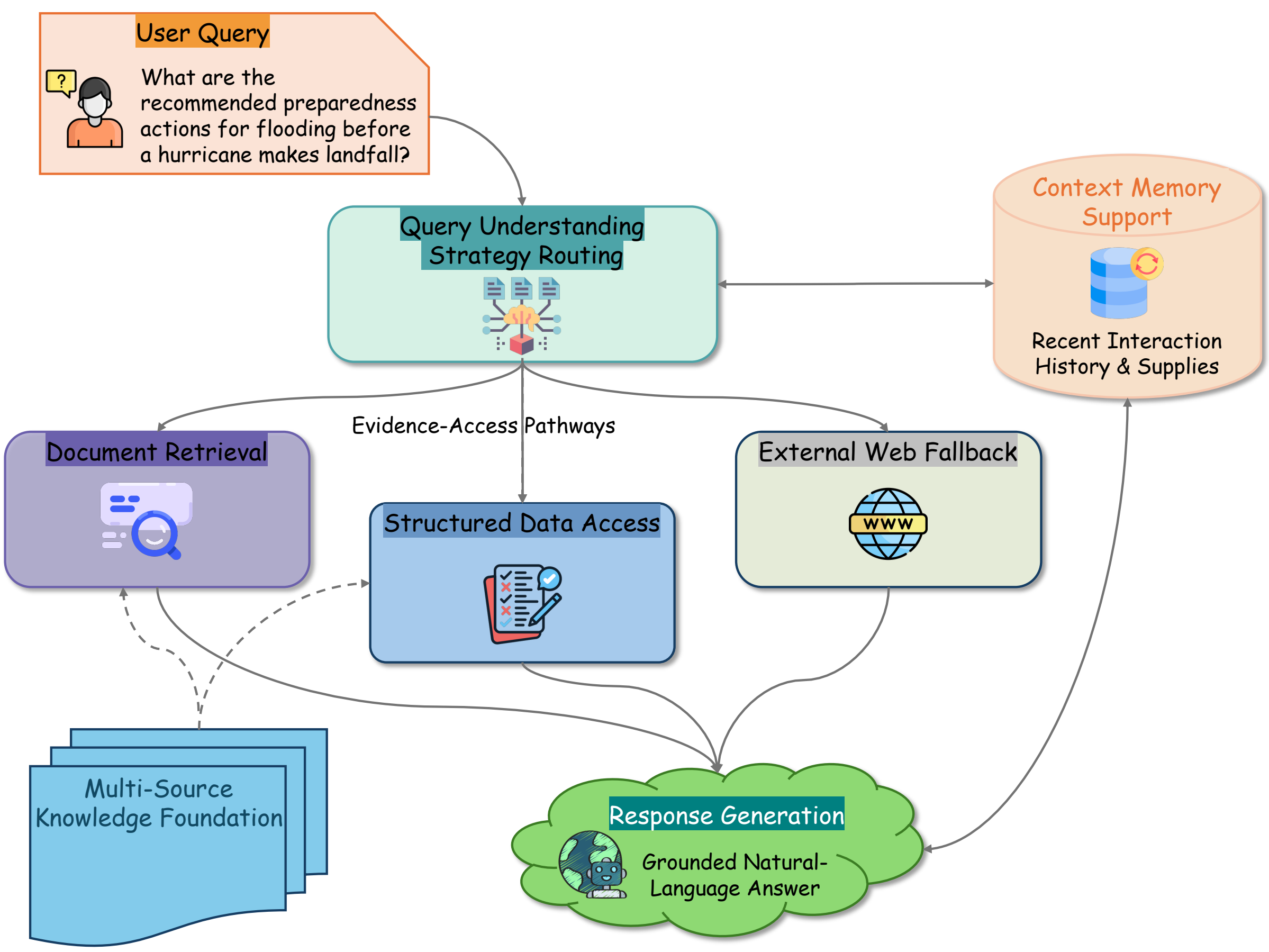

**Figure 1.** Schematic overview of the proposed DisastRAG system.

### 3.2 Multi-Source Knowledge Foundation

The system's knowledge foundation consists of two layers that reflect the heterogeneous nature of disaster information: a structured layer for quantitative records and an unstructured layer for document-based knowledge. These layers provide the evidence basis for supporting different types of disaster information requests within a unified framework.

The structured layer contains disaster-related tabular records organized in a relational schema. It covers Hurricanes Harvey (2017) and Beryl (2024) and includes multiple record types, such as power outage counts, rainfall and stream flooding indicators, evacuation rates derived from mobility data, building damage assessments, and preparedness-related activity counts for essential commercial infrastructure, including gas stations, grocery stores, pharmacies, and home improvement stores. These records are georeferenced at ZIP-code and census-tract levels and support information needs that depend on exact values, temporal comparisons, and geographic aggregation.

The unstructured layer consists of hazard-related documents collected from national, regional, and institutional sources spanning disaster preparedness, response, recovery, and policy across multiple hazard types, including hurricanes, floods, storm surge, wildfire, extreme heat, and drought. Source materials include federal and state policy documents, agency technical reports, academic publications, and online resources from organizations such as FEMA, NOAA, ATSDR, PreventionWeb, and regional emergency management agencies. To support downstream evidence access, documents are segmented into passage-level units and paired with source identifiers.

### 3.3 Query Understanding and Strategy Routing

The LLM-informed query understanding and strategy routing layer serves as the system's coordination mechanism for request-dependent evidence access. It consists of two coordinated components: a query understanding module, which transforms the incoming request into a structured intermediate representation, and a strategy router, which uses that representation to dispatch the request to the appropriate evidence-access pathway (Figure 2).

Each incoming user request is first processed by the query understanding module, which operates in a GPT-assisted mode and constructs a structured query representation through three sequential LLM-based steps. First, the module performs conversational query rewriting to resolve coreferences and implicit references. Up to three recent interaction turns are retrieved from the session memory bank, and GPT-4o is instructed, with temperature set to 0.3 and a maximum output length of 100 tokens, to generate a fully self-contained rewritten query that replaces ambiguous expressions such as "it," "this event," or "those counties" with explicit entities derived from prior context. This step ensures that downstream processing operates on an explicit and context-complete query. Second, the rewritten query is analyzed by GPT-4o to infer the query type, detect ambiguity, and determine domain relevance. Query types are categorized as quantitative, descriptive, explanatory, contextual, or other. Quantitative queries seek field-indexed, numerical, or aggregative information; descriptive queries request definitions, procedural guidance, or factual summaries; explanatory queries seek causal or interpretive understanding; locational queries emphasize geographic scope or place-specific conditions; contextual queries depend on prior conversational references for interpretation; and other captures residual cases that do not fit these categories. Third, an entity-tag extraction step identifies disaster-type and location references in the query and returns them as structured tags. Together, the rewritten query, inferred query type, ambiguity flag, domain-relevance indicator, and extracted entity tags form the structured query representation, which serves as a normalized intermediate form for downstream routing and retrieval.

The strategy router then uses this structured query representation to determine the most appropriate evidence-access pathway. Routing decisions are made based on the inferred query type. Descriptive and explanatory queries, including requests for procedural guidance, hazard interpretation, and document-based situational context, are routed to the document retrieval branch. Quantitative queries involving field-indexed or aggregative information, such as outage counts, shelter capacity, or evacuation statistics, are routed to the structured data retrieval branch. Queries marked as out-of-domain bypass internal retrieval and are instead directed to the external web fallback.

The extracted entity tags further support this process by reinforcing disaster- and location-consistent evidence and by restricting memory-bank retrieval to geographically relevant prior context when conversational history is incorporated. Through this design, the LLM-informed query understanding and strategy routing layer transforms conversational input into a structured, context-aware representation that dynamically coordinates evidence access across heterogeneous disaster information sources.

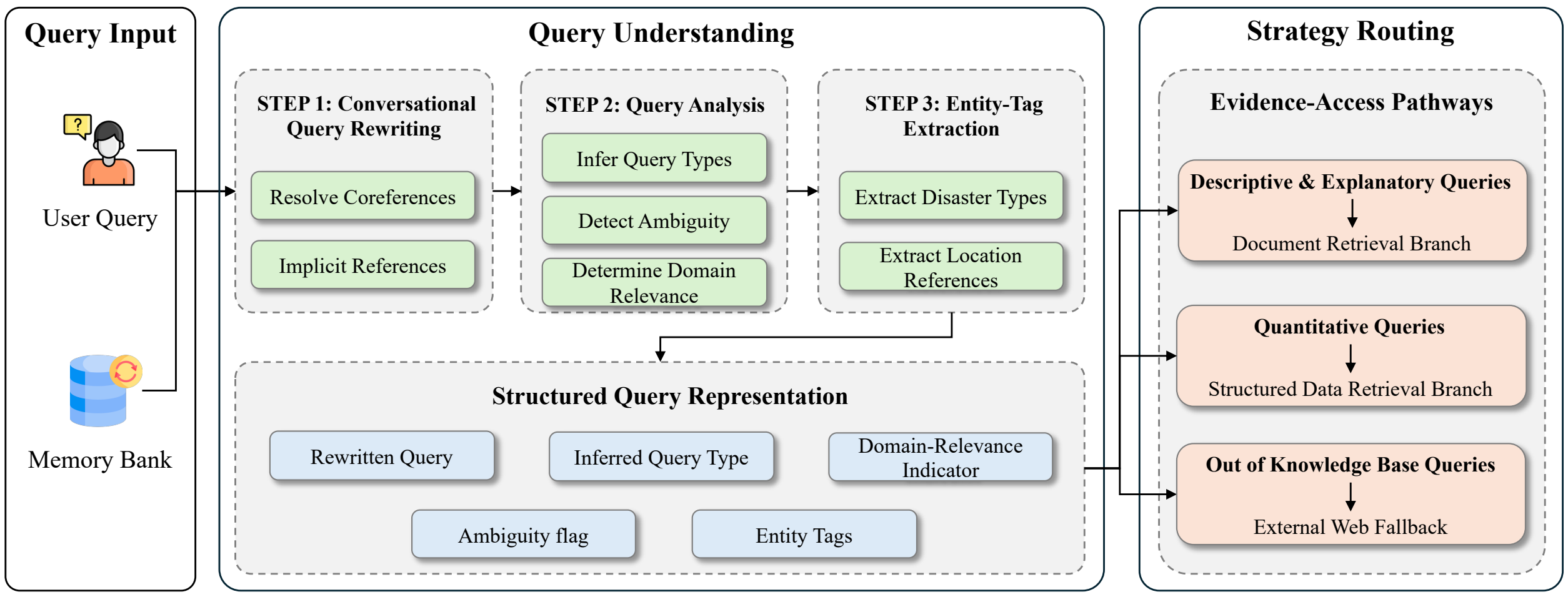

**Figure 2.** Architecture diagram of query understanding and strategy routing modules

### 3.4 Evidence-Access Pathways

To address the heterogeneous information needs of disaster management, the evidence-access layer is organized around three parallel pathways that route user requests to the most appropriate source of evidence (presented in Figure 3). The three pathways enable the system to align evidence access with the informational characteristics of the user query.

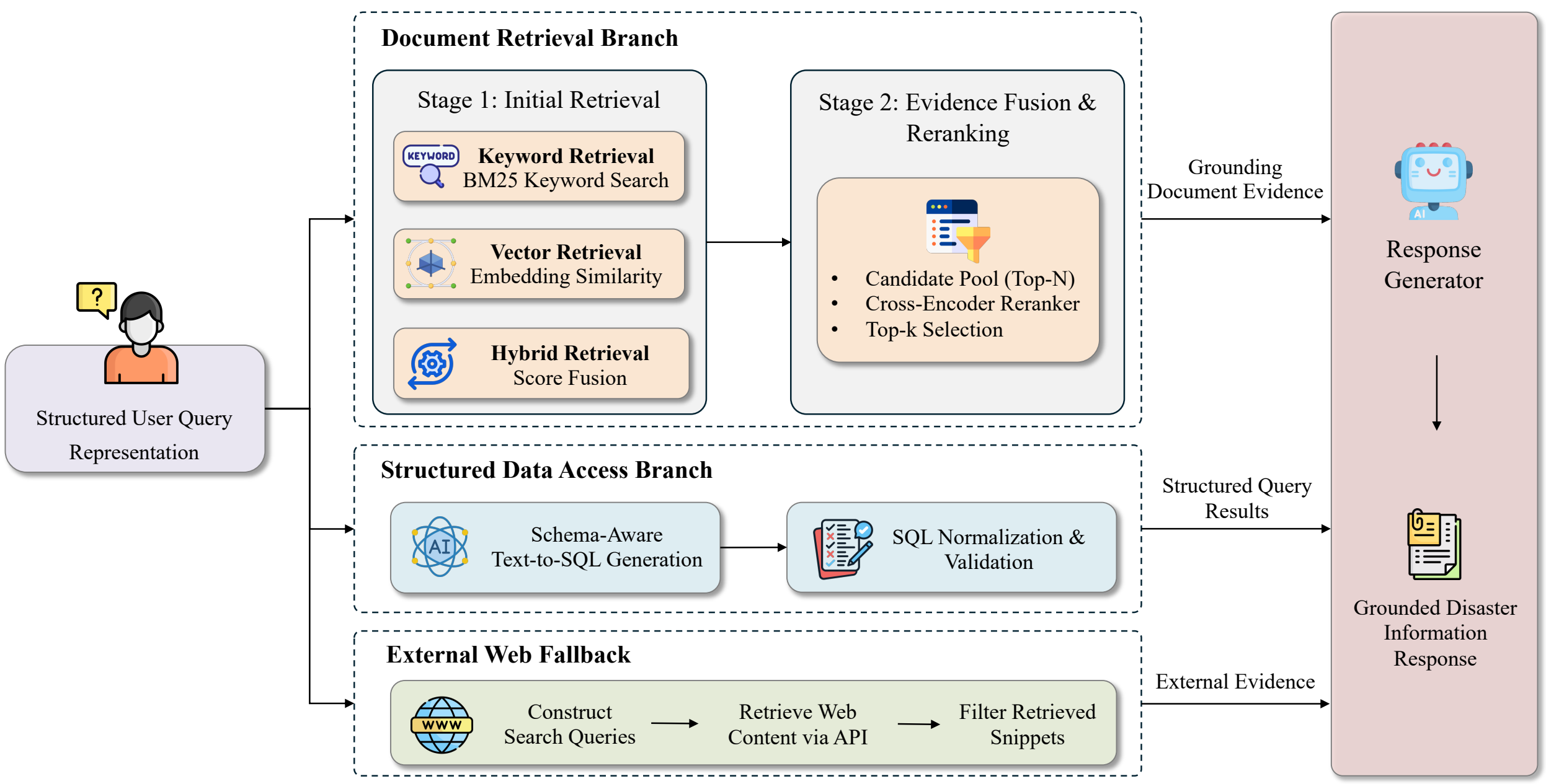

**Figure 3.** Evidence-access pathways in the proposed DisastRAG system.

### 3.4.1 Document Retrieval Branch

The document pathway handles descriptive and explanatory requests that are best answered from the unstructured corpus. It follows a two-stage design to support evidence grounding for document-centered queries. In the first stage, the system retrieves an initial candidate set of passages relevant to the user request. In the second stage, these candidates are further refined through evidence fusion and reranking to produce the final grounding context for response generation.

The initial retrieval module implements three retrieval strategies: keyword-based, vector-based, and hybrid, which are systematically evaluated across configurations in Section 5. Keyword-based retrieval applies lexical matching over a preconstructed BM25 index to identify passages with direct term overlap with the user query (Robertson and Zaragoza 2009). This strategy is effective for disaster-domain requests containing explicit terms that appear directly in the corpus, such as disaster names, geographic identifiers, event-specific terminology, and other domain-relevant vocabulary. In such cases, exact lexical correspondence may be more important than broader semantic similarity. Among the keyword retrieval variants examined in this study, Elasticsearch was selected as the lexical retrieval backend based on its strongest retrieval performance (Elasticsearch 2018). Vector-based retrieval operates over the chunk-level unstructured corpus, in which each passage is encoded offline as a dense embedding and stored in a vector index. At inference time, the user request is encoded into the same embedding space using the intfloat/e5-small-v2 bi-encoder model (Wang, Yang et al. 2022), and relevant passages are retrieved through approximate nearest-neighbor search over the indexed embeddings. This strategy captures semantic similarity between the request and candidate passages and is useful for queries that paraphrase, generalize, or otherwise differ from the wording used in the source documents. Hybrid retrieval combines the candidate outputs of the keyword and vector channels into a unified ranked pool. Specifically, the scores from each channel are independently normalized by their respective maximum values and then linearly combined with equal weights to produce a merged relevance score. The resulting ranked pool integrates lexical precision with semantic recall to provide broader and more balanced evidence coverage.

The evidence fusion and reranking stage serves as the refinement layer of the document retrieval branch. Its role is to aggregate candidate passages returned from the retrieval module, construct a unified candidate pool, and select the final evidence set used for grounded response generation. Following retrieval, the system constructs a candidate pool from the top-N ranked passages returned by the active retrieval channel. The candidate pool size (IR ∈ {100, 150, 200}) is treated as a configurable design parameter and is systematically evaluated in Section 5. When hybrid retrieval is active, the candidate pool is formed from the merged outputs of the keyword and vector retrieval channels, after which duplicate passages are removed and the pool is sorted prior to reranking. Then the candidate pool is reranked using the cross-encoder model cross-encoder/ms-marco-MiniLM-L-6-v2 (Bajaj, Campos et al. 2016), which jointly encodes the query and each candidate passage to produce fine-grained relevance scores. Based on these scores, the reranker selects the final top-k passages (Rerank ∈ {5, 10, 15}) to serve as the grounding evidence set for the response generator.

In this way, retrieval and reranking play complementary roles within the framework: retrieval determines the breadth of the candidate evidence, whereas reranking determines the precision of the final evidence passed to generation. Section 5 systematically evaluates the effects of both design choices on downstream performance.

### 3.4.2 Structured Data Access Branch

The structured access is designed for requests whose answers depend on exact values or structured comparisons. Its role is to provide database-grounded evidence for queries that cannot be answered reliably through document retrieval alone. When a request is routed to this branch, the system translates the natural-language query into an executable SQL statement using an LLM-guided Text-to-SQL procedure. Query generation is performed using schema-aware prompting, in which GPT-4o is instructed to construct SQL statements using the available tables, columns, and join relationships defined in the database schema. The prompt also includes domain-specific guidance that maps common disaster-related expressions to corresponding SQL operations, such as aggregation, ranking, and geographic grouping. This design constrains the generation process and reduces the likelihood of invalid queries or schema mismatches. As illustrated in Figure 4, the prompt template consists of two components: task-level instructions and schema-aware context. The instruction block defines the SQL generation task and provides procedural guidance for translating natural-language disaster queries into executable SQL. The schema and domain guidance block provides the relational schema, available join keys, and example mappings between natural-language expressions and SQL operators.

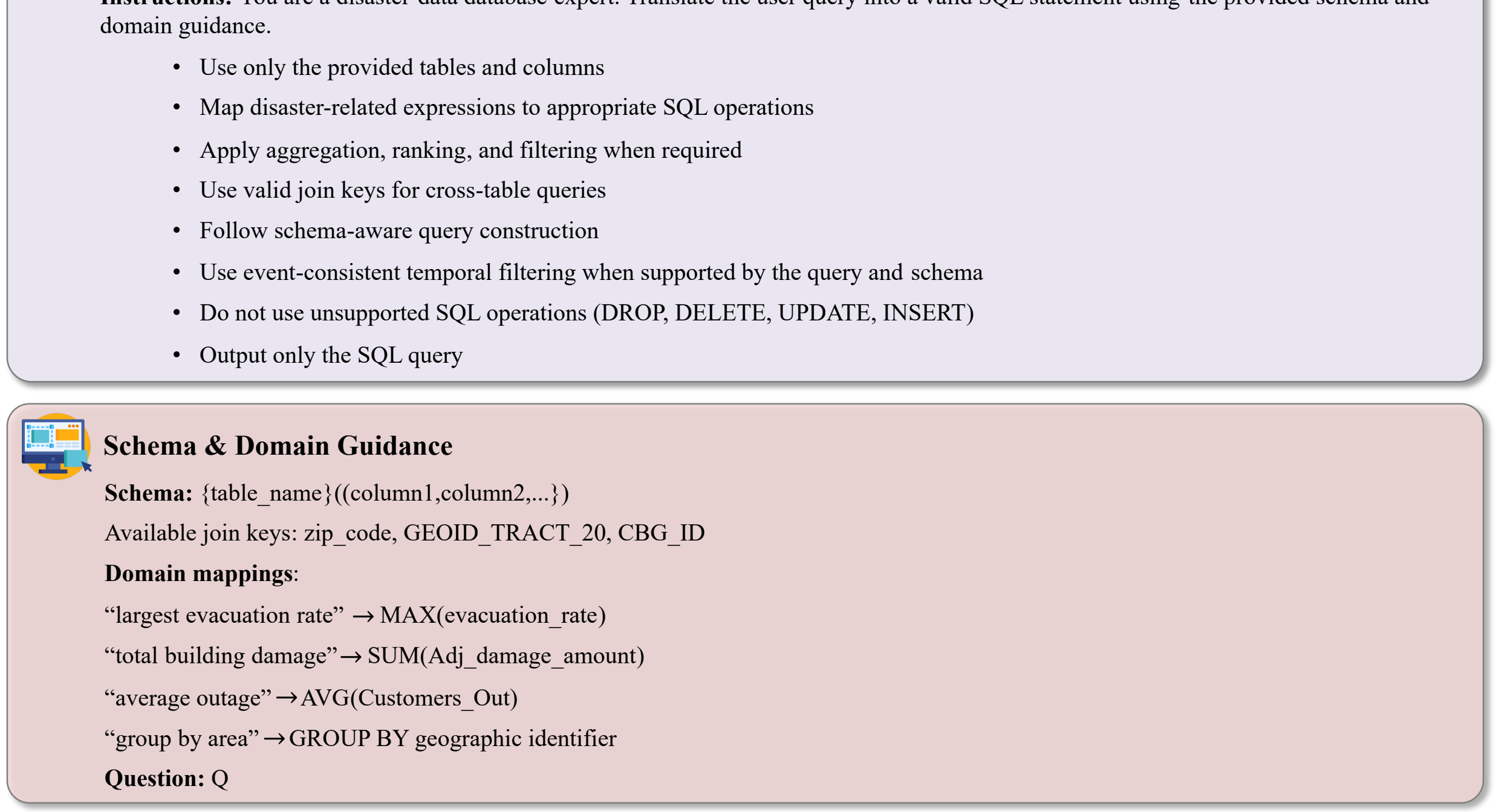

**Figure 4.** Prompt template for schema-aware Text-to-SQL generation in the structured data access branch.

Following generation, the produced SQL query is passed through a lightweight normalization and validation stage. This step standardizes column references, checks consistency with the database schema, and filters unsupported operations prior to execution. Queries that cannot be translated into valid executable SQL are redirected to the external web fallback branch. This fallback mechanism preserves system robustness when structured access is not feasible.

### 3.4.3 External Web Fallback

The external web fallback supports requests whose information requirements cannot be sufficiently satisfied by the internal knowledge foundation. Its role is to extend the framework beyond the internal corpus by formulating targeted external search queries, retrieving relevant web-based evidence, and filtering that evidence before it is incorporated into downstream response generation. For a routed external request, the module constructs a search query from the structured request representation produced during query understanding, using extracted attributes such as the referenced disaster event, geographic scope, temporal context, and informational intent. Relevant web content is then retrieved through external search API DuckDuckGo (Hands 2012). Because external results may vary in quality and relevance, retrieved snippets are filtered with respect to the active request context, with particular attention to spatial and temporal consistency. The retained evidence is subsequently assembled into the grounding context used by the response generation module.

## 3.5 Response Generation

The response generation module transforms the evidence assembled by the selected access pathway into a grounded natural-language response. This module is designed to accommodate heterogeneous inputs produced by different branches of the system. Prior to generation, the framework constructs a response context from the evidence returned by the selected pathway. For document-centered requests, this context is formed from the top reranked textual passages; for structured information requests, it is derived from executed SQL query results formatted as structured rows; and for external fallback cases, it is assembled from filtered web-retrieved snippets. Relevant prior turns from the session memory bank may also be incorporated to preserve conversational continuity and support interpretation of follow-up requests.

As illustrated in Figure 5, the framework uses branch-specific prompt templates to align response generation with the informational form of the available evidence. Prompting in the document retrieval branch emphasizes grounded synthesis from retrieved passage. In the structured data access branch, prompting is oriented toward translating structured query results into human-readable summaries while preserving fidelity to exact numerical values and ranked comparisons. In the external web fallback branch, prompting places greater emphasis on selective synthesis, preference for authoritative sources, and explicit acknowledgment of uncertainty. In this way, response generation serves as the stage where heterogeneous evidence is consolidated, interpreted, and translated into an interactive form for disaster information access.

**Document Retrieval Prompt**

**Task**

Generate a grounded answer using retrieved disaster-related documents.

**Input**

- Conversation history
- Retrieved document chunks
- User query

**Instruction**

- Synthesize across retrieved passages
- Maintain concise factual tone
- Prioritize recent and relevant information
- Acknowledge missing information

**Output/Constraint**

Structured natural language response grounded in retrieved context.

**Structured Data Access Prompt**

**Task**

Generate a natural language summary from structured query results.

**Input**

- Generated SQL query
- Top-k rows formatted as structured records

**Instruction**

- Summarize key patterns and rankings
- Interpret results in disaster context
- Use exact numeric values
- Avoid technical database terminology

**Output/Constraint**

Summary, Key findings, Top results, Context

Do not mention SQL or database operations.

**External Search Prompt**

**Task**

Generate a disaster-relevant answer using filtered web search snippets.

**Input**

- Conversation context
- Web search snippets

**Instruction**

- Prioritize official sources
- Synthesize across multiple snippets
- Include statistics when available
- Focus on disaster-relevant information
- Acknowledge uncertainty

**Output/Constraint**

Do not include URLs in answer text.

**Figure 5.** Prompt template for branch-specific response generation.

### 3.6 Contextual Memory Support

The contextual memory component is a supporting module that extend its ability to operate in more realistic disaster information environments. At each interaction turn, the memory bank stores a structured entry containing the user query, the generated answer, extracted entity tags, and a timestamp. These entries are maintained in a fixed-length sliding window so that the session context remains bounded and reflects the immediate interaction history.

Memory retrieval follows a two-stage priority mechanism. First, the system searches prior entries for those whose stored entity tags match the current query's disaster and location context, allowing partial string matches to accommodate abbreviated references. If such entries are found, the most relevant matching turns are returned; otherwise, the system falls back to the most recent turns in the session history. Retrieved entries are formatted as ordered question–answer pairs and supplied to the relevant downstream component.

The retrieved memory is incorporated where prior interaction context is most useful. It helps the query understanding module interpret follow-up questions that contain implicit or abbreviated references, and it provides the response generation stage with relevant conversational background so that answers remain coherent across turns. When external web evidence is used, memory also helps preserve continuity with the ongoing discussion. In this way, this module serves as a lightweight mechanism for maintaining interaction continuity within the broader multi-source framework.

# 4. Experimental Design

## 4.1 Evaluation Scope

The study adopts a two-part evaluation design that reflects the different roles of components within the overall framework. The primary quantitative focus is the retrieval and reranking pipeline, which serves as the core evidence-grounding mechanism of the document retrieval branch. Specifically, the evaluation compares keyword-based, vector-based, and hybrid retrieval strategies, examines candidate-pool size and reranking depth as configurable design factors, and measures how these choices affect downstream performance on representative information-seeking tasks. The remaining components of the broader system are examined through representative case studies in Section 6, which illustrate how the full architecture supports heterogeneous information requests.

## 4.2 Evaluation Task Design

The evaluation tasks follow the Human–LLM collaboration framework introduced in DisastQA (Chen, Yin et al. 2026), which constructs evidence-grounded disaster-response questions from query–passage pairs. Tasks are organized into two formats. Multiple-choice (MCQ) items evaluate evidence-supported factual identification by selecting the correct answer among plausible distractors, while open-ended (OE) items assess explanatory and synthesis-oriented information needs through free-form responses grounded in retrieved passages. These tasks are used to represent a range of disaster information access needs, spanning factual identification, evidence-supported selection, explanation, and synthesis across disaster-related contexts.

### 4.3 Compared Settings and Baselines

The experimental evaluation compares settings along two axes: retrieval-side configuration and LLM backbone. A no-retrieval condition is included for each model as a baseline.

#### 4.3.1 Retrieval Configurations

Three initial retrieval strategies are evaluated: keyword-based retrieval, vector-based retrieval, and hybrid retrieval. For each retrieval strategy, nine RAG configurations are evaluated by crossing three initial candidate pool sizes (IR $\in$ {100, 150, 200}) with three reranking depths (Rerank $\in$ {5, 10, 15}), yielding a $3 \times 9 = 27$ configuration design. The pool size determines how many passages are retrieved before reranking; the reranking depth determines how many of the reranked passages are passed to the generator. Reranking is performed over the full candidate pool in a single batch pass (batch size 128). For MCQ tasks, retrieved passages are concatenated into a grounding context capped at 6,000 tokens. For OE tasks, each passage is independently truncated to a maximum of 512 tokens before context assembly.

### 4.3.2 Baseline

A no-retrieval (i.e., pure generation) baseline is established for each model, in which the LLM receives only the question text with no augmented context. This condition measures intrinsic parametric knowledge of the disaster events and provides the reference against which all retrieval augmentation gains are quantified.

### 4.3.3. Model Backbones

Four open-source LLMs are evaluated without task-specific fine-tuning: DeepSeek-V3, LLaMA-3-8B, Mistral-7B, and Qwen-3-8B. All models are applied using greedy decoding for MCQ tasks. For OE tasks, generation uses temperature 0.7 with difficulty-adaptive maximum token budgets: 80 tokens (easy), 180 tokens (medium), 300 tokens (hard), and 400 tokens (extremely complex). This design isolates differences in how models of varying parameter scale and instruction-following capacity utilize the same retrieved evidence context.

## 4.4 Evaluation Metrics

### 4.4.1 MCQ Accuracy

For multiple-choice tasks, performance is measured using accuracy over $N$questions to reflect closed-form response quality. Let $y_i \in \{A, B, C, D\}$ denote the gold answer for item $i$, and $\hat{y}_i$ the predicted answer. MCQ accuracy is defined as Equation 1:

$$Acc_{MCQ} = \frac{1}{N}\sum_{i=1}^{N} 1[\hat{y}_i = y_i] \tag{1}$$

This metric evaluates evidence-supported factual discrimination under a controlled response format.

### 4.4.2 OE Keypoint Coverage

For open-ended tasks, performance is measured using Keypoint Coverage, which evaluates factual completeness of generated responses. For each item, a set of gold keypoints derived from the human reference answer is used as the evaluation target, and keypoints are extracted from the model-generated response using GPT-4o for semantic comparison. Coverage is then computed as the proportion of gold keypoints that are semantically supported by the generated response. Let $K_i^*$ denote the gold keypoint set for item $i$, with cardinality $m_i$, and let $e\left(k_{i,j}, \hat{K}_i\right) \in \{0, 1\}$ indicate whether gold keypoint $k_{i,j}$ is semantically supported by the generated response. Per-item coverage is defined as Equation 2:

$$Cov(\hat{r}_i, K_i^*) = \frac{1}{m_i}\sum_{j=1}^{m_i} e(k_{i,j}, \hat{K}_i) \tag{2}$$

The final OE score is the mean coverage across all $N$ items:

$$\overline{Cov} = \frac{1}{N}\sum_{i=1}^{N} Cov(\widehat{r_i}, K_i^*) \quad (3)$$

# 5. Quantitative Results

## 5.1 MCQ Performance Across Retrieval Configurations

Figure 6 presents MCQ accuracy under multiple retrieval configurations for four LLM backbones. The three subplots correspond to vector, hybrid, and keyword retrieval, respectively, and dashed horizontal lines indicate the corresponding no-retrieval baseline for each model. Table 1 summarizes the best-performing retrieval configuration for each backbone. Across the evaluated configurations, retrieval augmentation consistently improves MCQ accuracy relative to the no-retrieval condition for all four models. This pattern indicates that incorporating retrieved evidence reliably enhances evidence-grounded factual discrimination in disaster information retrieval, while also demonstrating that these gains remain robust across different retrieval mechanisms.

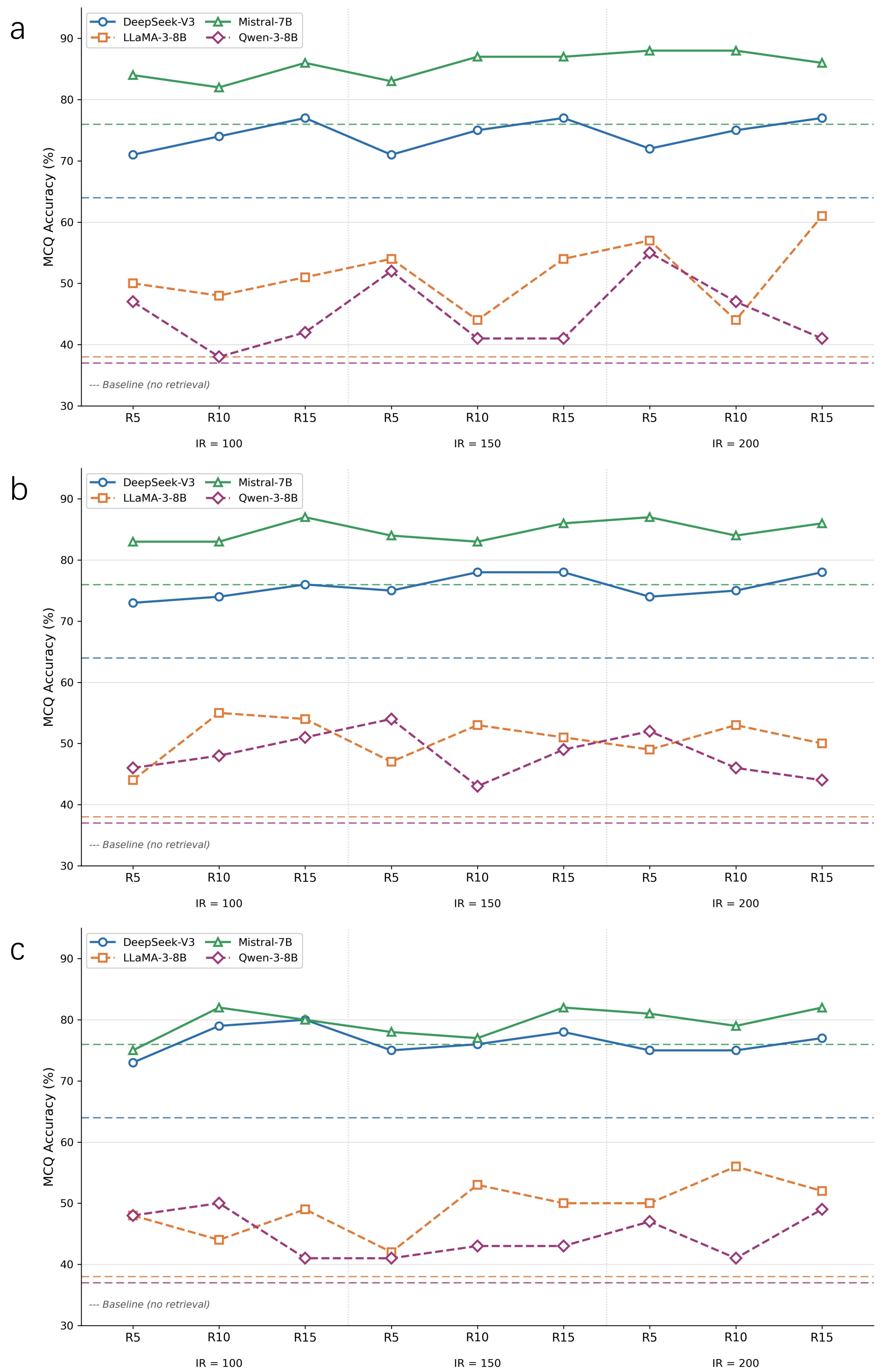
a
b
c
DeepSeek-V3
Mistral-7B
LLaMA-3-8B
Qwen-3-8B
MCQ Accuracy (%)
--- Baseline (no retrieval)
R5
R10
R15
IR = 100
IR = 150
IR = 200
30
40
50
60
70
80
90

**Figure 6.** MCQ accuracy under three different retrieval strategies across nine configurations: (a) vector-based retrieval, (b) keyword-based retrieval, and (c) hybrid retrieval. Dashed lines indicate no-retrieval baselines.

Mistral-7B achieves the highest absolute MCQ accuracy among the four evaluated backbones, reaching 88.0% under vector retrieval at IR200-R5 and IR200-R10, which represents an improvement of 12.0 percentage points over its 76.0% no-retrieval baseline. Its performance is comparatively stable across retrieval methods and configurations, with only modest variation between its strongest and weakest settings, suggesting that a stronger backbone is less sensitive to retrieval-side design choices. DeepSeek-V3 achieves its highest MCQ accuracy of 80.0% under hybrid retrieval at IR100-R15, representing an improvement of 16.0 percentage points over its 64.0% baseline. DeepSeek-V3 appears to benefit more from the combination of dense and sparse retrieval signals than from either strategy in isolation, as both vector and keyword retrieval peak only in the 77–78% range. This pattern suggests that for DeepSeek-V3, lexical overlap provides useful complementary evidence alongside semantic retrieval for evidence-supported answer selection. LLaMA-3-8B and Qwen-3-8B begin with the weakest no-retrieval baselines (38.0% and 37.0%, respectively) and exhibit the largest absolute gains from retrieval augmentation, improving by 23.0 and 18.0 percentage points under their best configurations.

LLaMA-3-8B reaches its highest MCQ accuracy of 61.0% under vector retrieval at IR200-R15, while Qwen-3-8B peaks at 55.0% under vector retrieval at IR200-R5. At the same time, both models show substantially greater sensitivity to retrieval configuration than the stronger backbones, with performance varying markedly across the nine tested settings. This pattern indicates that retrieval can compensate meaningfully for weaker baseline disaster-domain performance, but that the resulting gains are highly configuration-dependent and therefore require careful tuning in practical deployment.

Across retrieval configurations, increasing the candidate pool size from IR = 100 to IR = 200 yields modest and model-dependent improvements. The performance of stronger backbones (Mistral-7B and DeepSeek-V3) remains relatively stable across IR values, while weaker models (LLaMA-3-8B and Qwen-3-8B) show larger fluctuations, with their best results often appearing at IR = 200. This suggests that weaker models benefit more from a larger candidate pool, which increases the likelihood of retrieving relevant evidence. Similarly, moving from R = 5 to R = 15 produces only minor changes for Mistral-7B and DeepSeek-V3, indicating that the top-ranked candidates are already of high quality. By contrast, LLaMA-3-8B and Qwen-3-8B exhibit greater sensitivity to reranking depth, but without a consistent pattern across retrieval strategies. These results indicate that deeper reranking yields limited additional gains when relevant evidence is already captured within the top-ranked candidates.

**Table 1.** MCQ accuracy summary for four LLM backbones under best retrieval Configurations

| Model | Base Accuracy | Best RAG Accuracy | Best Configuration | Absolute Gain |
|---|---|---|---|---|
| Mistral-7B | 76.0% | 88.0% | Vector Retrieval IR200-Rerank5/Rerank10 | +12.0 pp |
| DeepSeek-V3 | 64.0% | 80.0% | Hybrid Retrieval IR100-Rerank15 | +16.0 pp |
| LLaMA-3-8B | 38.0% | 61.0% | Vector Retrieval IR200-Rerank15 | +23.0 pp |
| Qwen-3-8B | 37.0% | 55.0% | Vector Retrieval IR200-Rerank5 | +18.0 pp |

### 5.2 Open-Ended Performance Across Retrieval Configurations

To examine retrieval effects on open-ended factual completeness, Figure 7 presents OE keypoint coverage across four LLM backbones and multiple retrieval settings. The three subplots correspond to vector hybrid, and keyword retrieval, while dashed lines denote the no-retrieval baselines. The best-performing retrieval configuration for each backbone is reported in Table 2. Compared with MCQ, improvements in OE performance are more moderate across all models, reflecting the greater challenge of producing free-text responses that fully capture the reference keypoints.

Even so, retrieval augmentation consistently improves OE coverage over the no-retrieval baseline for each model, with gains ranging from 7.2 to 10.5 percentage points.

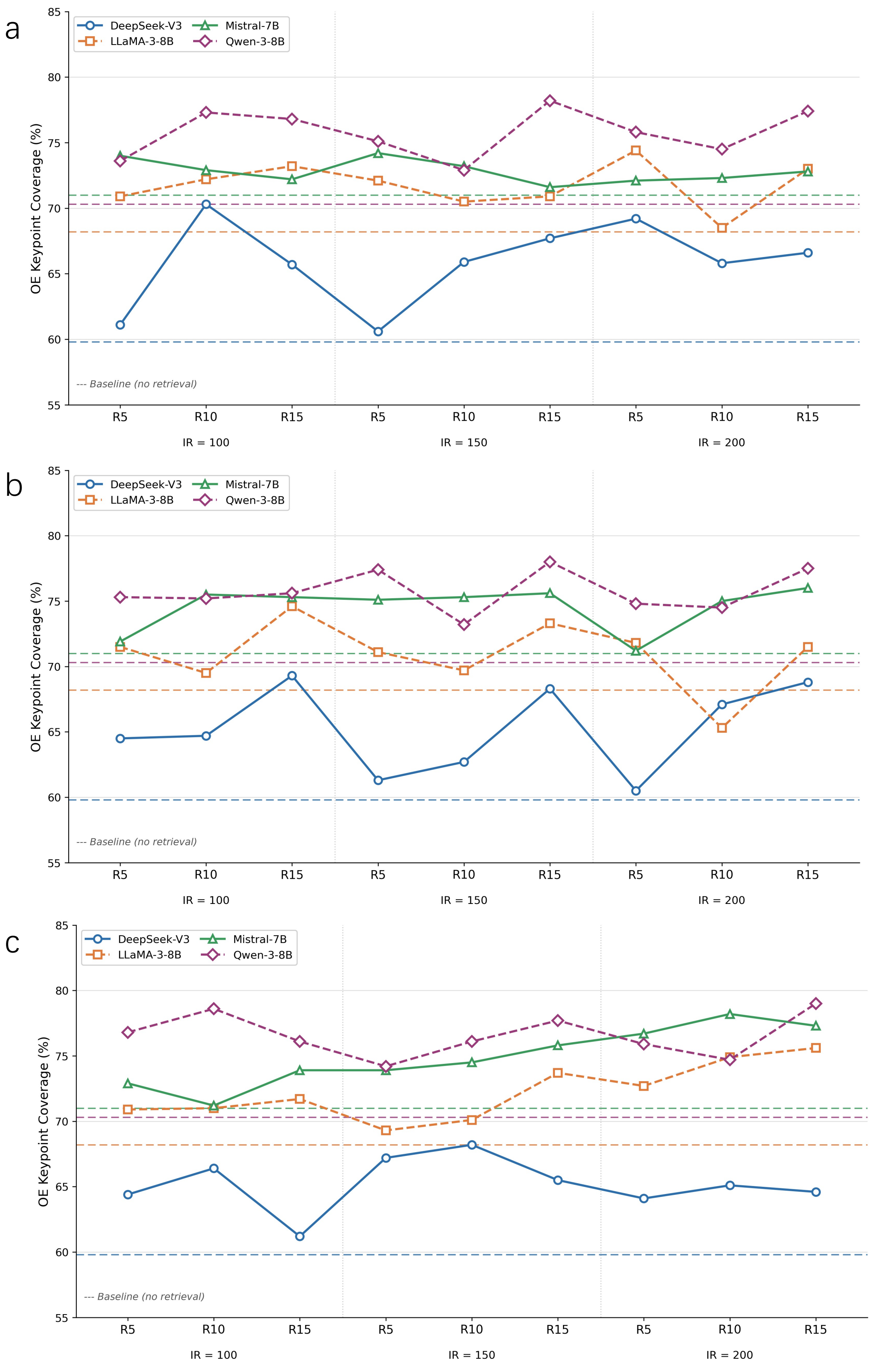
a
b
c
DeepSeek-V3
Mistral-7B
LLaMA-3-8B
Qwen-3-8B
OE Keypoint Coverage (%)
--- Baseline (no retrieval)
R5
R10
R15
IR = 100
IR = 150
IR = 200

**Figure 7.** OE keypoint coverage under three different retrieval strategies across nine configurations: (a) vector-based retrieval, (b) keyword-based retrieval, and (c) hybrid retrieval. Dashed lines indicate no-retrieval baselines.

DeepSeek-V3 exhibits the largest absolute gain, rising from 59.8% at baseline to 70.3% under vector retrieval at IR100-R10. This pattern suggests that when a model's open-ended baseline is relatively weak, semantically retrieved evidence can substantially improve factual completeness by supplying missing answer elements that are not reliably available from parametric knowledge alone. LLaMA-3-8B, Qwen-3-8B, and Mistral-7B begin with comparatively stronger OE baselines (68.2%, 70.3%, and 71.0% respectively) and show smaller but consistent improvements under retrieval augmentation. Among them, Qwen-3-8B achieves the highest absolute OE coverage, reaching 79.0% under hybrid retrieval at IR200-R15. Mistral-7B achieves its best OE result of 78.2% under hybrid retrieval at IR150-R5, while LLaMA-3-8B peaks at 75.6% under hybrid retrieval at IR200-R5. The strongest configurations for these models tend to occur at larger candidate pools (IR150–200), indicating that retaining a broader set of potentially relevant passages can help the model cover multiple keypoints required for complete answers.

Most models reach their best performance under hybrid retrieval, suggesting that combining lexical and semantic signals is generally more effective for improving response completeness in open-ended disaster questions. Although vector retrieval remains competitive for DeepSeek-V3, it is outperformed by hybrid retrieval for the other three backbones. Keyword retrieval generally performs below both vector and hybrid retrieval across models and configurations, indicating that semantic matching is especially important when complete answers require integrating multiple relevant information points.

**Table 2.** OE keypoint coverage summary for four LLM backbones under best retrieval configurations

| Model | Base Coverage | Best RAG Coverage | Best Configuration | Absolute Gain |
|---|---|---|---|---|
| DeepSeek-V3 | 59.8% | 70.3% | Vector Retrieval IR100-Rerank10 | +10.5 pp |
| Qwen-3-8B | 70.3% | 79.0% | Hybrid Retrieval IR200-Rerank15 | +8.7 pp |
| LLaMA-3-8B | 68.2% | 75.6% | Hybrid Retrieval IR200-Rerank5 | +7.4 pp |
| Mistral-7B | 71.0% | 78.2% | Hybrid Retrieval IR150-Rerank5 | +7.2 pp |

## 6. Case Studies of System-Oriented Disaster Information Access

### 6.1 Case Design and Presentation Format

The quantitative evaluation in Section 5 focuses on the document retrieval branch, while the broader architecture also includes additional evidence-access pathways designed for request types that the document retrieval pipeline alone cannot adequately address. This section illustrates the two other pathways through representative case studies: a structured access interface for quantitative, field-indexed requests over relational disaster records, and an external web fallback for requests that require information beyond the coverage of the internal corpus.

The structured access case shows how the system translates a natural-language request into a database query and returns precise record-level evidence. The external fallback case shows how the system extends beyond its internal knowledge foundation to retrieve methodological or forward-looking information from external sources. These cases illustrate the framework's query-adaptive routing behavior and show that its evidence-access architecture is not limited to a single fixed retrieval pipeline.

### 6.2 Structured Data Access Pathway

This case illustrates how the framework handles a record-oriented, quantitative information request through the structured data access pathway. Figure 8 displays the screenshot the DisastRAG system for this interaction. The user asks: "Which area has the largest evacuation rate during Hurricane Harvey? I want to know it in zip code level". The query understanding module identifies this as a quantitative query seeking field-indexed and aggregative information from structured records. Entity-tag extraction further identifies the disaster type as Hurricane Harvey and the requested geographic reporting unit as zip code level. This structured query representation is passed to the strategy router, which dispatches the request to the structured access branch.

Within the structured access branch, the Text-to-SQL procedure translates the natural-language request into an executable SQL statement over the harvey_evacuation_data table. The query groups records by zip code and sorts them by maximum evacuation rate in descending order. The executed SQL query returns a ranked list of zip codes and their associated maximum evacuation rates. Among the returned results, zip code 77061 is identified as the area with the highest observed evacuation rate (57.14%), followed by 77025 and 77005, each with 55.56%. These structured results are passed to the response generation module as quantitative evidence, from which the system produces a natural-language summary highlighting both the highest-ranked location and the broader variation across the top-ranked zip codes.

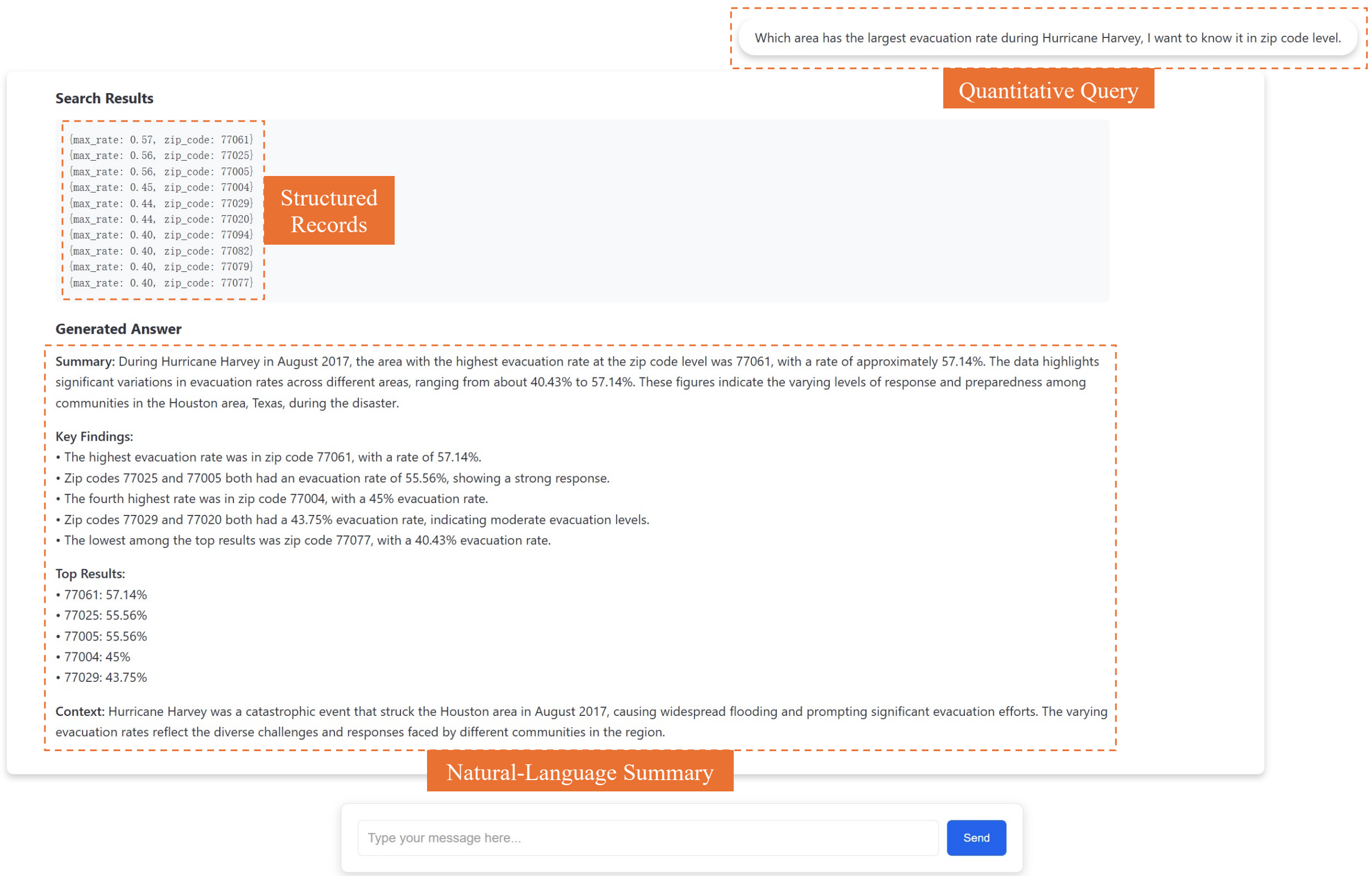

**Figure 8.** Structured access pathway in action: the system translates a natural-language evacuation-rate query into SQL over Hurricane Harvey records and returns a ranked zip-code-level response.

### 6.3 External Web Fallback Pathway

This case illustrates how the framework extends beyond its internal knowledge foundation through the external web fallback pathway. Figure 9 shows the system screenshot of this interaction. The user asks: “I want to predict the flood inundation depth of the road network in Houston assuming there will be a hurricane similar to Harvey”. The query understanding module determines this request falls outside the types of questions covered by the internal pathways, and the strategy router therefore dispatches it to the external web fallback branch. In the system interface, this routing decision is made explicit through a provenance label indicating that the response is grounded in web-retrieved evidence.

Within the external web fallback branch, the system formulates a targeted search query from the structured request representation produced during query understanding. This query is submitted to the DuckDuckGo search API, which returns a set of snippets related to remote sensing, hydrological simulation, and flood inundation mapping. The retrieved snippets are then filtered for relevance to the active request context, with priority given to sources that are spatially consistent with Houston and the Gulf Coast region, while snippets referring to unrelated geographies or events are excluded. The retained evidence is assembled into the grounding context for response generation. Based on this context, the generation module synthesizes a response that identifies three key methodological elements for flood-inundation prediction. The final response further emphasizes that reliable prediction depends on current geographic and meteorological inputs and that the unusual characteristics of Hurricane Harvey may limit the direct transferability of Harvey-based assumptions to future events.

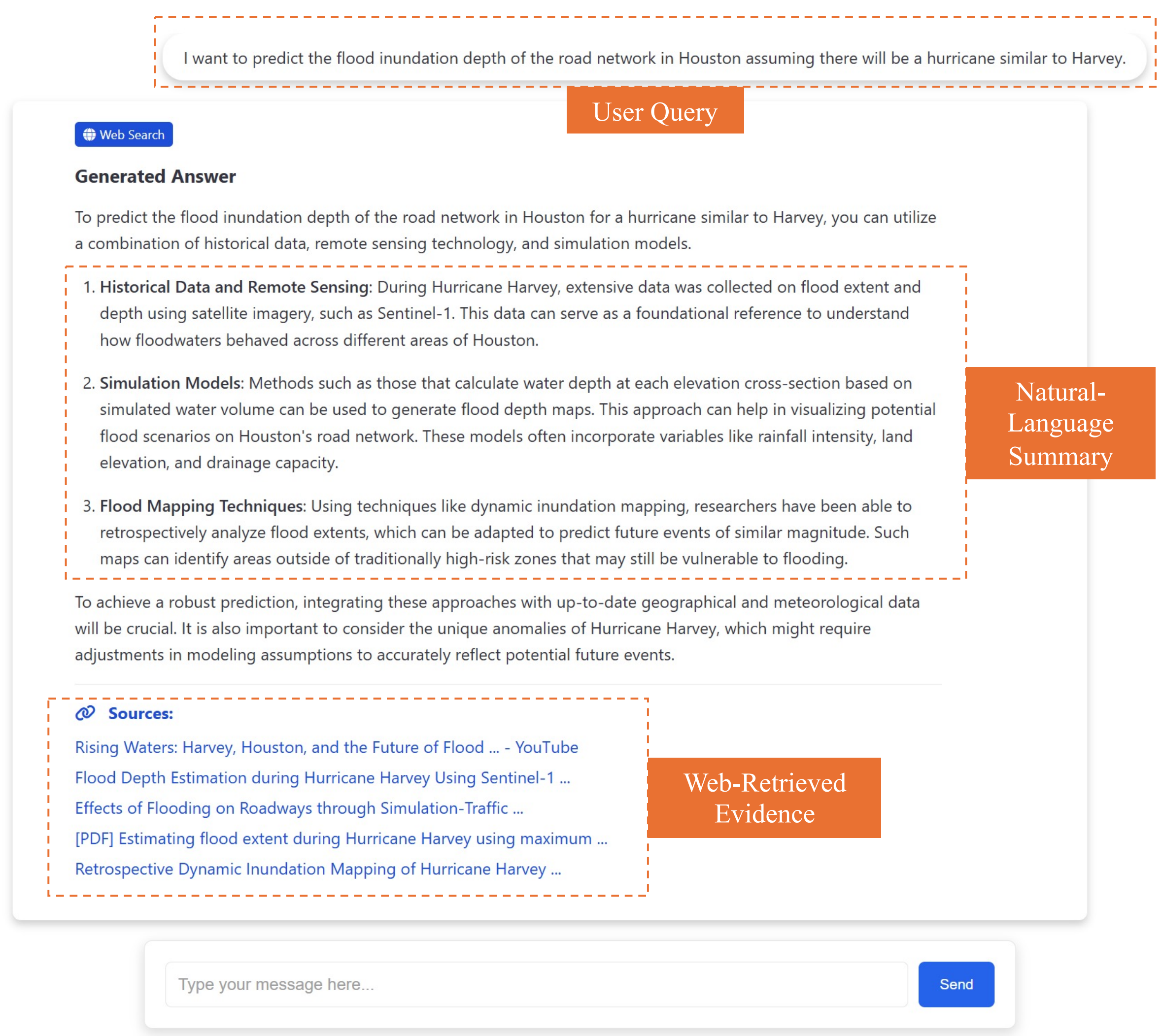

**Figure 9.** External fallback pathway in action: the system routes a predictive flood-modeling query to web search and synthesizes methodological guidance from retrieved external sources

## 7. Discussion and Concluding Remarks

This paper presented DisastRAG, a disaster-oriented framework for information integration and access built on an LLM-powered, retrieval-augmented architecture.

This study aims to address a disaster management challenge that relevant information is fragmented across heterogeneous sources, and remains difficult to access and coordinate in a unified way. In response, the framework organizes disaster information access through three complementary pathways: document-centered retrieval over a curated hazard corpus, structured query access over relational disaster records, and an external web fallback for requests that exceed the coverage of the internal knowledge base. By identifying the informational characteristics of each user request, the system routes it to the appropriate evidence-access mode and generates responses grounded in the corresponding evidence source.

Within this broader architecture, this study focused its quantitative analysis on the primary evidence-grounding mechanism of the document retrieval branch. Across four open-source LLM backbones and two task formats, the results showed that retrieval augmentation consistently improves performance, while the effects of candidate-pool size and reranking depth vary across models and task types. The case studies further illustrate that the system's capabilities extend beyond document retrieval alone. In particular, structured access supports exact record-level requests over relational disaster data, whereas external fallback supports requests requiring predictive or methodological information unavailable in the internal corpus.

In this study, we frame disaster information access as a multi-source coordination problem. In this respect, the study shows how natural-language interaction, pathway selection, and heterogeneous evidence access can be integrated together within a unified system, while also clarifying how retrieval and reranking design shape downstream response quality. In this way, this study contributes both practical guidance for configuring retrieval pipelines in disaster-domain RAG systems and a broader system-oriented view of how structured, unstructured, and external evidence can be coordinated to support a wider range of disaster information needs.

This study has limitations. The quantitative evaluation focuses primarily on the document retrieval branch, while the structured access and external fallback pathways are illustrated through representative case studies. In addition, routing accuracy, multi-turn interaction quality, and real-time pathway selection are not directly evaluated. Future work can address these limitations by developing larger-scale evaluations for the additional pathways, assessing routing and memory-supported interaction more directly, and testing the framework in more realistic human-in-the-loop or practitioner-facing environments. Extending the system to broader disaster datasets, more diverse request types, and dynamic external information sources would further strengthen its applicability as a coordinated system for disaster information integration and access.

## Data availability

All datasets used in this study are derived from publicly available sources.

## Code availability

The code that supports the findings of this study is available from the corresponding author upon request.

## Acknowledgements

This work was supported by the National Science Foundation under Grant CMMI-1846069 (CAREER). Any opinions, findings, conclusions, or recommendations expressed in this research are those of the authors and do not necessarily reflect the view of the funding agency.

## Competing interests

The authors declare no competing interests.